# A New Model for Pricing Collateralized Financial Derivatives

Tim Xiao[1]

Risk Models, Capital Markets, BMO, Toronto, Canada

## ABSTRACT

This paper presents a new model for pricing financial derivatives subject to collateralization. It allows for collateral arrangements adhering to bankruptcy laws. As such, the model can back out the market price of a collateralized contract. This framework is very useful for valuing outstanding derivatives. Using a unique dataset, we find empirical evidence that credit risk alone is not overly important in determining credit-related spreads. Only accounting for both collateral posting and credit risk can sufficiently explain unsecured credit costs. This finding suggests that failure to properly account for collateralization may result in significant mispricing of derivatives. We also empirically gauge the impact of collateral agreements on risk measurements. Our findings indicate that there are important interactions between market and credit risk.

Collateral arrangements are regulated by the Credit Support Annex (CSA) and are always counterparty-based as different counterparties may have different CSA agreements. Thus, financial institutions normally group derivatives into counterparty portfolios first and then process them separately. The difference between counterparties is determined by counterparty credit qualities whereas the difference in collateralization is distinguished by the terms and conditions of CSA agreements.

Collateralization is a critical component of the plumbing of the financial system. The use of collateral in financial markets has increased sharply over the past decade, yet analytical and empirical research on collateralization is relatively sparse. The effect of collateralization on valuation and risk is an understudied area.

Due to the complexity of collateralization, the literature seems to turn away from direct and detailed modeling. For example, Johannes and Sundaresan [2007], and Fuijii and Takahahsi [2012] model collateralization via a cost-of-collateral instantaneous rate. Piterbarg [2010] regards collateral as a regular asset and uses the replication approach to price collateralized derivatives.

Contrary to previous studies, we present a model that characterizes a collateral process directly based on the fundamental principal and legal structure of CSA. The model is devised that allows for collateralization adhering to bankruptcy laws. As such, it can back out price changes due to counterparty risk and collateral posting. Our model is very useful for valuing off-the-run or outstanding derivatives.

This article makes theoretical and empirical contributions to the study of collateralization by addressing several essential questions. First, how does collateralization affect swap rate?

Interest rate swaps collectively account for two-thirds of all outstanding derivatives (see FinPricing). An ISDA mid-market swap rate is based on a mid-day polling. Dealers use this market rate as a reference and make some adjustments to quote an actual swap rate. The adjustment or swap premium is determined by many factors, such as credit risk, liquidity risk, funding cost, operational cost and expected profit, etc.

Unlike generic mid-market swap rates, swap premia are determined in a competitive market according to the basic principles of supply and demand. A swap client first contacts a number of swap dealers for a quotation and then chooses the most competitive one. If a premium is too low, the dealer may lose money. If a premium is too high, the dealer may lose the competitive advantage.

Unfortunately, we do not know the detailed allocation of a swap premium, i.e., what percentage of the adjustment is charged for each factor. Thus, a direct empirical assessment of the impact of collateralization on swap rate is impossible.

To circumvent this difficulty, this article uses an indirect empirical approach. We define a *swap premium spread* as the premium difference between two swap contracts that have exactly the same terms and conditions but are traded with different CSA counterparties. We reasonably believe that if two contracts are identical except counterparties, the swap premium spread should reflect counterparty credit risk only, as all other risks/costs are identical.

Empirically, we obtain a unique proprietary dataset from an investment bank. We use these data and a statistical measurement $R^2$ to examine whether credit risk and collateralization, alone or in combination, are sufficient to explain market swap premium spreads. We first study the marginal impact of credit risk. Since credit default swap (CDS) premium theoretically reflects the credit risk of a firm, we use the market swap premium spreads as the response variable and the CDS premium differences between two counterparties as the explanatory variable. The estimation result shows that the adjusted $R^2$ is 0.7472, implying that approximately 75% of market spreads can be explained by counterparty credit risk. In other words, counterparty risk alone can provide a good but not overwhelming prediction on spreads.

We then assess the joint effect. Because implied or model-generated spreads take into account both counterparty risk and collateralization, we assign the model-implied spreads as the explanatory variable and the market spreads as the response variable. The new adjusted $R^2$ is 0.9906, suggesting that counterparty risk and collateralization together have high explanatory power on premium spreads. The finding leads to practical implications, such as collateralization modeling allows forecasting credit spread.

Second, how does collateralization affect counterparty credit risk? Credit value adjustment (CVA) is the most prominent measurement in counterparty credit risk. We select all the CSA counterparty portfolios in the dataset and then compute their CVAs. We find that the CVA of a collateralized counterparty portfolio is always smaller than the one of the same portfolio without collateralization. We also find that credit risk is negatively correlated with collateralization as an increase in collateralization causes a decrease in credit risk. The empirical tests corroborate our theoretical conclusions that collateralization can reduce CVA charges and mitigate counterparty risk.

Finally, how do collateralization and credit risk, either alone or in combination, impact market risk? How do they interact with each other? Value at risk (VaR) is the regulatory measurement for market risk We compute VaR in three different cases – VaR without taking credit risk into account, VaR with credit risk, and VaR with both credit risk and collateralization. We find that there is a positive correlation between market risk and credit risk as VaR increases after considering counterparty credit risk. We also find that collateralization and market risk have a negative correlation, i.e., collateral posting can actually reduce VaR. This finding contradicts the prevailing belief in the market that collateralization would increase market risk (see Collateral Management – Wikipedia).

The rest of this article is organized as follows: First we present a new model for pricing collateralized financial derivatives. Then we discuss empirical evidences. Finally, the conclusions and discussion are provided. All proofs and detailed derivations are contained in the appendices.

## Pricing Collateralized Financial Derivatives

A CSA is a legal document that regulates collateral posting. It specifies a variety of terms including threshold, independent amount, and minimum transfer amount (MTA). A threshold is the unsecured credit exposure that a party is willing to bear. A MTA is used to avoid the workload associated with a frequent transfer of insignificant collateral amounts. An independent amount plays the same role as an initial margin or haircut. We define the effective collateral threshold as the threshold plus the MTA. Collateral is called as soon as the mark-to-market (MTM) value rises above the effective threshold.

There are three types of collateralization: partial, over or full. A positive effective threshold corresponds to partial-collateralization where the posting of collateral is less than the MTM value. A negative effective threshold represents over-collateralization where the posting of collateral is greater than the MTM value. A zero-value effective threshold equates with full-collateralization where the posting of collateral is equal to the MTM value. Our generic model is applicable to all the types.

Since the only reason for taking collateral is to reduce/eliminate credit risk, collateral analysis should be closely related to credit risk modeling. There are two primary types of models that attempt to describe default processes in the literature: structural models and reduced-form models. Many practitioners in the market have tended to gravitate toward the reduced-from models given their mathematical tractability and market consistency.

We consider a filtered probability space $(\Omega, \mathcal{F}, \{\mathcal{F}_t\}_{t \geq 0}, \mathcal{P})$ satisfying the usual conditions, where $\Omega$ denotes a sample space, $\mathcal{F}$ denotes a $\sigma$-algebra, $\mathcal{P}$ denotes a probability measure, and $\{\mathcal{F}_t\}_{t \geq 0}$ denotes a filtration. In the reduced-form framework, the stopping or default time $\tau$ of a firm is modeled as a Cox arrival process whose first jump occurs at default and is defined by,

$$\tau = \inf\left\{t : \int_0^t h(s, \Gamma_s) ds \geq \Delta\right\} \tag{1}$$

where $h(t)$ or $h(t, \Gamma_t)$ denotes the stochastic hazard rate dependent on an exogenous common state $\Gamma_t$, and $\Delta$ is a unit exponential random variable independent of $\Gamma_t$.

It is well-known that the survival probability from time $t$ to $s$ in this framework is defined by

$$p(t,s) := P(\tau > s \mid \tau > t) = \exp\left(-\int_t^s h(u) du\right) \tag{2a}$$

The default probability for the period $(t, s)$ is given by

$$q(t,s) := P(\tau \leq s \mid \tau > t) = 1 - p(t,s) = 1 - \exp\left(-\int_t^s h(u) du\right) \tag{2b}$$

Let valuation date be *t*. Consider a financial contract that promises to pay a $X_T > 0$ at maturity date $T > t$, and nothing before the time.

The binomial default rule considers only two possible states: default or survival. For a discrete one-payment period (*t*, *T*) economy, at time $T$ the contract either defaults with the default probability $q(t,T)$ or survives with the survival probability $p(t,T)$. The survival payoff is equal to $X_T$ and the default payoff is a fraction of $X_T$: $\varphi X_T$, where $\varphi$ is the recovery rate. The value of this defaultable contract at time *t* is the discounted expectation of all the possible payoffs and is given by

$$V^N(t) = E\{D(t,T)[p(t,T) + \varphi(T)q(t,T)]X_T | \mathcal{F}_t\} = E[D(t,T)I(t,T)X_T | \mathcal{F}_t] \qquad (3)$$

where $E\{\bullet|\mathcal{F}_t\}$ is the expectation conditional on $\mathcal{F}_t$, $D(t,T)$ denotes the risk-free discount factor at time *t* for maturity *T* and $I(t,T) = [p(t,T) + \varphi(T)q(t,T)]$ can be regarded as a risk-adjusted discount ratio.

Suppose that there is a CSA agreement between a bank and a counterparty in which the counterparty is required to deliver collateral when the mark-to-market (MTM) value arises over the effective threshold *H*.

The choice of modeling assumptions for collateralization should be based on the legal structure of CSA. According to the Bankruptcy Law, if the demand for default payment exceeds the collateral value, the balance of the demand will be treated as an unsecured claim and subject to its pro rata distribution under the Bankruptcy Code's priority scheme (see Garlson [1991], Routh and Douglas [2005], and Edwards and Morrison [2005]). The default payment under a CSA can be mathematically expressed as

$$P^D(T) = C(T) + \varphi(T)(X_T - C(T)) = \varphi(T)X_T + C(T)(1 - \varphi(T)) \qquad (4)$$

where $C(T)$ is the collateral amount at *T*.

It is worth noting that the default payment in equation (4) is always greater than the original recovery, i.e., $P^D(T) > \varphi(T)X_T$ because $\varphi(T)$ is always less than 1. Said differently, the default payoff of a collateralized contract is always greater than the default payoff of the same contract without a CSA.

That is why the major benefit of collateralization should be viewed as an improved recovery in the event of a default.

According to CSA, if the contract value $V^C(t)$ is less than the effective threshold $H(t)$, no collateral is posted; otherwise, the required collateral is equal to the difference between the contract value and the effective threshold. The collateral amount posted at time *t* can be expressed mathematically as

$$C(t) = \max(V^C(t) - H(t), 0) \tag{5}$$

where $H(t) = 0$ corresponds to full-collateralization; $H(t) > 0$ represents partial-collateralization; and $H(t) < 0$ reflects over-collateralization.

For a discrete one-payment period (*t, T*) economy, at time *T*, if the contract survives, the survival value is the promised payoff $X_T$ and the collateral taker returns the collateral to the provider. If the contract defaults, the default payment is defined in (4) where the future value of the collateral is $C(T) = C(t)/D(t,T)$. Since the most predominant form of collateral is cash according to ISDA [2013], it is reasonable to consider the time value of money only for collateral assets. The large use of cash means that collateral is both liquid and not subject to large fluctuations in value. The above collateral rule tells us that collateral does not have any bearing on survival payoffs; instead, it takes effect on default payments only. The value of the CSA contract is the discounted expectation of all the payoffs and is given by

$$V^C(t) = E\{D(t,T)[q(t,T)(C(T) + \varphi(T)(X_T - C(T))) + p(t,T)X_T] | \mathcal{F}_t\} \tag{6}$$

After some simple mathematics, we have the following proposition

***Proposition 1***: *The value of a collateralized single-payment contract is given by*

$$V^C(t) = E[F(t,T)X_T | \mathcal{F}_t] - G(t,T) \tag{7a}$$

where

$$F(t,T) = \left(1_{V^N(t) \leq H(t)} + 1_{V^N(t) > H(t)} / \bar{I}(t,T)\right) I(t,T) D(t,T) \tag{7b}$$

$$G(t,T) = 1_{V^N(t) > H(t)} H(t) \bar{q}(t,T)(1 - \varphi(T)) / \bar{I}(t,T) \tag{7c}$$

where $\bar{I}(t,T) = E(I(t,T)|\mathcal{F}_t)$. $I(t,T)$ and $V^N(t)$ are defined in (3).

Proof: See the Appendix.

We may think of $F(t,T)$ as the CSA-adjusted discount factor and $G(t,T)$ as the cost of bearing unsecured credit risk. Proposition 1 tells us that the value of a collateralized contract is equal to the present value of the payoff discounted by the CSA-adjusted discount factor minus the cost of taking unsecured counterparty credit risk. This proposition theoretically demonstrates that collateral posting changes valuation.

The pricing in Proposition 1 is relatively straightforward. We first compute $V^N(t)$ and then test whether its value is greater than $H(t)$. After that, the calculations of $F(t,T)$, $G(t,T)$ and $V^C(t)$ are easily obtained.

We discuss a special case where $H(t) = 0$ corresponding to full-collateralization. Suppose that default probabilities are uncorrelated with interest rates and payoffs[2]. From Proposition 1, we can easily obtain $V^C(t) = V^F(t)$ where $V^F(t) = E[D(t,T)X_T|\mathcal{F}_t]$ is the risk-free value. That is to say: the value of a fully-collateralized contract is equal to the risk-free value. This conclusion is in line with the results of Johannes and Sundaresan [2007], Fuijii and Takahahsi [2012], and Piterbarg [2010].

Proposition 1 can be easily extended from single payment to multiple payments. Suppose that a defaultable contract has $m$ cash flows represented as $X_i$ with payment dates $T_i$, where $i = 1,…,m$. We derive the following proposition:

***Proposition 2:*** *The value of a collateralized multiple-payment contract is given by*

$$V^C(t) = \sum_{i=1}^{m} E\left[\prod_{j=0}^{i-1}(F(T_j,T_{j+1}))X_i \Big| \mathcal{F}_t\right] - \sum_{i=0}^{m-1} E\left[\prod_{j=0}^{i-1}(F(T_j,T_{j+1}))G(T_i,T_{i+1}) \Big| \mathcal{F}_t\right] \qquad (8a)$$

---

[2] Moody's Investor's Service [2000] presents statistics that suggest that the correlations between interest rates, default probabilities and recovery rates are very small and provides a reasonable comfort level for the uncorrelated assumption.

where

$$F(T_j, T_{j+1}) = \left(1_{J(T_j,T_{j+1}) \leq H(T_j)} + 1_{J(T_j,T_{j+1}) > H(T_j)} / \bar{I}(T_j, T_{j+1})\right) I(T_j, T_{j+1}) D(T_j, T_{j+1}) \qquad (8b)$$

$$G(T_j, T_{j+1}) = 1_{J(T_j,T_{j+1}) > H(T_j)} H(T_j) \bar{q}(T_j, T_{j+1})(1 - \varphi(T_{j+1})) / \bar{I}(T_j, T_{j+1}) \qquad (8c)$$

$$J(T_j, T_{j+1}) = E\left[D(T_j, T_{j+1}) I(T_j, T_{j+1})(V^C(T_{j+1}) + X_{j+1}) \middle| \mathcal{F}_{T_j}\right] \qquad (8d)$$

where $\prod_{j=0}^{i-1} F(T_j, T_{j+1}) = 1$ is an empty product when $i = 0$. Empty product allows for a much shorter mathematical presentation of many subjects.

The valuation in Proposition 2 has a backward nature. The intermediate values are vital to determine the final price. For a payment period, the current price has a dependence on the future price. Only on the final payment date $T_m$, the value of the contract and the maximum amount of information needed to determine $J(T_{m-1}, T_m)$, $F(T_{m-1}, T_m)$ and $G(T_{m-1}, T_m)$ are revealed. This type of problem can be best solved by working backward in time, with the later value feeding into the earlier ones, so that the process builds on itself in a recursive fashion, which is referred to as *backward induction*. The most popular backward induction algorithms are lattice/tree and regression-based Monte Carlo, i.e., Longstaff-Schwartz approach.

## Empirical Results

### Impact of collateralization on swap rate

In this subsection, we choose interest rate swaps for our empirical study. Ultimately, it is the objective of this subsection to test if counterparty credit risk and collateralization are sufficient to explain market swap premium spreads. We choose a statistical measurement $R^2$ to determine how much market spreads can be interpreted by model-implied spreads that take counterparty risk and collateralization into account.

Swap rate is the fixed rate that sets the market value of a swap at initiation to zero. ISDAFIX provides average mid-market swap rates based on a mid-day polling from a panel of dealers. In practice, the mid-market swap rates are generally not the actual swap rates transacted with counterparties, but are instead the benchmarks against which the actual swap rates are set. A swap dealer that arranges a contract and provides liquidity to the market involves costs. Therefore, it is necessary to adjust the mid-market swap rate to cover various transacting expenses and also to provide a profit margin to the dealer. As a result, the actual price agreed for a transaction is not zero but a positive amount to the dealer.

Unlike the generic benchmark swap rates, swap premia are determined according to the basic principles of supply and demand. The swap market is highly competitive. In a competitive market, prices are determined by the impersonal forces of demand and supply, but not by the manipulations of powerful buyers or sellers.

Prior research has primarily focused on the generic mid-market swap rates and results appear puzzling. Sorensen and Bollier [1994] believe that swap spreads are partially determined by counterparty default risk. Whereas Duffie and Huang [1996], Minton [1997] and Grinblatt [2001] find weak or no evidence of the impact of counterparty credit risk on swap spreads. Collin-Dufresne and Solnik [2001] and He [2001] further argue that many credit enhancement devices, e.g., collateralization, have essentially rendered swap contracts risk-free. Meanwhile, Duffie and Singleton [1999], and Liu, Longstaff and Mandell [2006] conclude that both credit and liquidity risks have an impact on swap spreads. Moreover, Feldhutter and Lando [2008] find that the liquidity factor is the largest component of swap spreads. It seems that there is no clear-cut answer yet regarding the relative contribution of various factors.

In contrast to previous research, this subsection mainly studies swap adjustments/premia related to credit risk and collateralization. It empirically measures the effect of collateralization on pricing and compares it with model-implied prices.

A swap premium is supposed to cover the expected profit and all the expenses, including the cost of bearing unsecured credit risk. Unfortunately, however, we do not know what percentage of the market swap premium is allocated to the unsecured credit risk, which makes a direct verification impossible.

To circumvent this difficulty, we design an indirect verification process in which we select some CSA swap pairs such that the two contracts in each pair have exactly the same terms and conditions but are traded with different counterparties under different collateral agreements. It is reasonable to believe that the difference between the two contracts in each pair is solely attributed to collateralized counterparty credit risk, as all the other risks/costs are identical. Therefore, by accounting for credit risk and collateralization, we can efficiently compare the implied spreads with the market spreads for these pairs.

We obtain a unique proprietary dataset from an investment bank. The dataset contains derivative contract data, counterparty data (including collateral agreements, recovery rates, etc), and market data. The trading dates are from May 6, 2005 to May 11, 2012. We find a total of 1002 swap pairs in the dataset, where the two contracts in each pair have the same terms and conditions but are traded with different CSA counterparties. We arbitrarily select one pair shown in Exhibit 1.

**Exhibit 1: A pair of 20-year swap contracts**

This exhibit displays the terms and conditions of two swap contracts that have different counterparties but are otherwise the same. We hide the counterparty names according to the security policy of the investment bank while everything else is authentic.

|  | Swap 1 | | Swap 2 | |
| --- | --- | --- | --- | --- |
|  | **Fixed leg** | **Floating leg** | **Fixed leg** | **Floating leg** |
| Counterparty | $X$ | | $Y$ | |
| Effective date | 15/09/2005 | 15/09/2005 | 15/09/2005 | 15/09/2005 |
| Maturity date | 15/09/2025 | 15/09/2025 | 15/09/2025 | 15/09/2025 |
| Day count | 30/360 | ACT/360 | 30/360 | ACT/360 |
| Payment frequency | Semi-annually | Quarterly | Semi-annually | Quarterly |
| Swap rate | **4.9042%** | - | **4.9053%** | - |
| Roll over | Mod_follow | Mod_follow | Mod_follow | Mod_follow |
| Principal | 25,000,000 | 25,000,000 | 25,000,000 | 25,000,000 |

| Currency | USD | USD | USD | USD |
| --- | --- | --- | --- | --- |
| Pay/receive | Bank receives | Party *X* receives | Bank receives | Party *Y* receives |
| Floating index | - | 3 month LIBOR | - | 3 month LIBOR |
| Floating spread | - | 0 | - | 0 |
| Floating reset | - | Quarterly | - | Quarterly |

An interest rate curve is the term structure of interest rates, derived from observed market instruments that represent the most liquid and dominant interest rate products for certain time horizons. Normally the curve is divided into three parts. The short end of the term structure is determined using LIBOR rates. The middle part of the curve is constructed using Eurodollar futures. The far end is derived using mid swap rates. The LIBOR-future-swap curve is presented in Exhibit 2. After bootstrapping the curve, we get the continuously compounded zero rates.

**Exhibit 2: USD LIBOR-future-swap curve**

This exhibit displays the closing mid prices as of September 15, 2005

| **Instrument Name** | **Price** |
| --- | --- |
| September 21 2005 LIBOR | 3.6067% |
| September 2005 Eurodollar 3 month | 96.1050 |
| December 2005 Eurodollar 3 month | 95.9100 |
| March 2006 Eurodollar 3 month | 95.8100 |
| June 2006 Eurodollar 3 month | 95.7500 |
| September 2006 Eurodollar 3 month | 95.7150 |
| December 2006 Eurodollar 3 month | 95.6800 |
| 2 year swap rate | 4.2778% |
| 3 year swap rate | 4.3327% |
| 4 year swap rate | 4.3770% |
| 5 year swap rate | 4.4213% |
| 6 year swap rate | 4.4679% |
| 7 year swap rate | 4.5120% |
| 8 year swap rate | 4.5561% |
| 9 year swap rate | 4.5952% |
| 10 year swap rate | 4.6368% |
| 12 year swap rate | 4.7089% |
| 15 year swap rate | 4.7957% |
| **20 year swap rate** | **4.8771%** |
| 25 year swap rate | 4.9135% |

As the payoffs of an interest rate swap are determined by interest rates, we need to model the evolution of floating rates. Interest rate models are based on evolving either short rates, instantaneous forward rates, or market forward rates. Since both short rates and instantaneous forward rates are not directly observable in the market, the models based on these rates have difficulties in expressing market views and quotes, and lack agreement with market valuation formulas for basic derivatives. On the other hand, the object modeled under the Libor Market Model (LMM) is market-observable. It is also consistent with the market standard approach for pricing caps/floors using Black's formula. They are generally considered to have more desirable theoretical calibration properties than short rate or instantaneous forward rate models. Therefore, we choose the LMM lattice proposed by Xiao [2011] for pricing collateralized swaps.

According to Proposition 2, we also need counterparty-related information, such as recovery rates, hazard rates and collateral thresholds. The CDS premia and recovery rates are given in Exhibit 3 and the collateral thresholds and MTAs of the CSA agreements are displayed in Exhibit 4. We can compute the hazard rates via a standard calibration process (see J.P. Morgan [2001]).

**Exhibit 3: CDS premia and recovery rates**

This exhibit displays the closing CDS premia as of September 15, 2005 and recovery rates

| Counterparty name | Bank | Company $X$ | Company $Y$ |
|---|---|---|---|
| 6 month CDS spread | 0.00031 | 0.000489 | 0.000808 |
| 1 year CDS spread | 0.000333 | 0.00056 | 0.001017 |
| 2 year CDS spread | 0.000516 | 0.000866 | 0.00154 |
| 3 year CDS spread | 0.000664 | 0.001147 | 0.002114 |
| 4 year CDS spread | 0.000848 | 0.00147 | 0.002768 |
| 5 year CDS spread | 0.001012 | 0.001783 | 0.003439 |
| 7 year CDS spread | 0.001334 | 0.002289 | 0.004283 |
| 10 year CDS spread | 0.001727 | 0.002952 | 0.005281 |
| 15 year CDS spread | 0.001907 | 0.003283 | 0.005814 |
| 20 year CDS spread | 0.002023 | 0.003266 | 0.006064 |
| 30 year CDS spread | 0.002021 | 0.00336 | 0.006461 |
| Recovery rate | 0.39213 | 0.35847 | 0.33872 |

**Exhibit 4: CSA agreement**

This exhibit provides the collateral thresholds and MTAs under the CSA agreements.

| CSA agreement | 1 | | 2 | |
|---|---|---|---|---|
| Counterparty name | Bank | Company $X$ | Bank | Company $Y$ |
| Threshold | 0 | 0 | 0 | 0 |
| MTA | 500000 | 500000 | 500000 | 500000 |

Given the above information, we are able to compute the collateralized swap rates. We first use the LMM to evolve the interest rates and then determine the associated CSA-adjusted discount factors as well as the cost of bearing unsecured credit risk according to Proposition 2. Finally, we calculate the collateralized swap rates via a backward induction method. The results are given in Exhibit 5.

**Exhibit 5: Swap rate results**

This exhibit presents the model-implied swap rates and premia as well as the dealer-quoted market swap rates and premia, where Swap premium (in bps) = Swap rate – Generic swap rate, and Premium spread = Premium of swap 2 – Premium of swap 1.

|  | Swap 1 | | Swap 2 | | Premium spread | Generic swap rate |
|---|---|---|---|---|---|---|
|  | Swap rate | Premium | Swap rate | Premium | | |
| Model-implied | 0.048780 | 0.09 bps | 0.048790 | 0.19 bps | 0.10 bps | 0.048771 |
| Dealer quoted | 0.049042 | 2.71 bps | 0.049053 | 2.82 bps | 0.11 bps | |

The 20-year generic mid-market swap rate is 0.048771 shown in Exhibit 2. The swap rates of contracts 1 and 2 are given in Exhibit 1 as 0.049042 and 0.049053. Accordingly, the market swap premia are 2.71 (0.049042 - 0.048771) basis points (bps) and 2.82 (0.049053 - 0.048771) bps. These premia are charged for many expenses. Although we do not know what percentage of the premia are allocated to cover the unsecured credit risks, we reasonably believe that the market premium spread, 0.11 (= 2.82 – 2.71) bps in Exhibit 5, should solely reflect the difference between the two counterparties' unsecured credit risks, as other factors are identical.

By accounting for both credit risk and collateralization, we calculate the model-implied swap rates as 0.048780 and 0.048790 shown in Exhibit 5. Consequently, the model-implied swap premia are 0.09 bps and 0.19 bps. The results imply that only a small portion of a swap premium is attributed to unsecured credit risk. Intuitively the small impact of credit risk and collateral posting on a swap premium is mainly due to 1) only the swap coupons rather than the national amount are exposed to counterparty risk; 2) the initial swap rate sets the contract value close to zero and there is only 50% chance to develop counterpart risk; 3) collateralization mitigates credit risk. This would certainly not be the case for other derivatives. The result is in line with the findings of Duffie and Huang [1996], Duffie and Singleton [1999], and Minton [1997].

Exhibit 5 shows that the model-implied spread is quite close to the dealer-quoted spread, suggesting that the model is fairly accurate in pricing collateralized financial instruments.

Repeating this exercise for the remaining pairs, we find that the model-implied spreads fluctuate randomly around the market spreads. We refer to the differences between the model-implied premium spreads and the market quoted premium spreads as the *model-market premium spread differentials*. The summary statistics of the market spreads, the model-implied spreads, and the model-market spread differentials are presented in Exhibit 6. It shows that the average of the model-market spread differentials is only -0.03 bps, which can be partly attributed to noise. The results indicate prima facie that the model performs quite well. The empirical tests corroborate the theoretical prediction on premium spreads.

**Exhibit 6: Summary statistics of model-implied swap premium spreads, market swap premium spreads and model-market swap premium spread differentials**

All values are displayed in bps. Model-market swap premium spread differential = Model-implied swap premium spread – Market-quoted swap premium spread.

|  | Max | Min | Mean | Median | Std |
|---|---|---|---|---|---|
| Market quoted swap premium spreads | 3.08 | -5.25 | -0.46 | -0.15 | 1.80 |
| Model-implied swap premium spreads | 2.10 | -5.33 | -0.44 | 0.03 | 1.73 |
| Model-market premium spread differentials | 0.99 | -1.18 | -0.03 | 0.05 | 0.46 |

Next, we examine the effects of credit risk and collateralization, alone or combined, on swap premium spreads. First, we study the marginal effect of credit risk. For each swap pair, we obtain the counterparty CDS premia. Presumably, the differences in CDS premia should mainly represent the differences in counterparty risk. To determine the strength of the statistical relationship between market premium spreads and counterparty risk, we present the estimate of the following regression model.

$$Y = a + bX + \varepsilon \tag{9}$$

where *Y* is the market swap premium spread, *X* is the difference between the counterparty CDS premia, *a* is the intercept, *b* is the slope, and $\varepsilon$ is the regression residual.

The results of this regression are shown in Exhibit 7. It can be seen that the adjusted $R^2$ is 0.7472, implying that approximately 74% of market premium spreads can be explained by credit risk alone. We provide empirical evidence that counterparty credit risk alone plays a significant but not overwhelming role in determining credit-related spreads when contracts are under collateral agreements. Moreover, the slope is 0.0127, suggesting that a CDS spread of about 100 bps translates into a swap spread of about 1.27 bps. Finally the small p-value indicates that the changes in CDS premia are closely related to the changes in market premium spreads.

**Exhibit 7: Marginal credit risk regression results**

This exhibit presents the regression results for the following regression model:

MarketSwapPremiumSpreads = a + b * DifferencesBetweenCounterpartyCDSs + $\varepsilon$

where the market swap premium spreads are used as the dependent variable and the differences between the counterparty CDS premia as the explanatory variable.

| Slope | Intercept | Adjusted $R^2$ | Significance F | T value | P value |
|---|---|---|---|---|---|
| 0.0127 | -2.7E-04 | 0.7472 | 1.1E-05 | 18.6 | 1.51E-66 |

According to ISDA Margin Survey (ISDA [2013], 73.7% of OTC derivatives are subject to collateral agreements. For large firms, the figure is 80.7%. Accounting for collateralization has become increasingly important in pricing OTC derivatives. Since the implied spreads generated by our model take into account both credit risk and collateralization, the statistical relationship between the market spreads and the model-implied spreads should refer to the joint effect of counterparty credit risk and collateralization on market spreads. Thus, we present another regression model where the market spreads are regressed on the implied spreads. The regression results are shown in Exhibit 8.

**Exhibit 8: Credit risk and collateralization combined regression results**

This exhibit presents the regression results for the following regression model:

$$\text{MarketSwapPremiumSpreads} = a + b * \text{ImpliedSwapPremiumSpreads} + \varepsilon$$

where the market swap premium spreads are used as the dependent variable and the model-implied swap premium spreads as the explanatory variable.

| Slope | Intercept | Adjusted $R^2$ | Significance F | T value | P value |
|---|---|---|---|---|---|
| 0.9857 | 4.48E-05 | 0.9906 | 3.21E-08 | 25.4 | 3.16E-110 |

Exhibit 8 shows that the adjusted $R^2$ value is 0.9906, implying that approximately 99% of the market spreads can be explained by the implied spreads. Also the small p-value suggests that the changes in implied spreads are strongly related to the changes in market spreads.

The empirical results shed light on the economic and statistical significance of collateralization. The increase in the explanatory power of swap premium spreads bears an interesting finding: It seems that credit risk alone has a modest explanatory power on premium spreads. Only the combination of credit risk and collateralization can sufficiently explain them.

**Impact of collateralization on credit risk**

In this subsection, we study how collateralization affects credit risk by measuring CVA changes due to collateral posting. CVA is the market price of counterparty credit risk that has become a central part of counterparty credit risk management.

From the same dataset above, we find that there are a total of 3052 counterparties having live trades as of May 11, 2012. 516 of them have CSA agreements. We randomly select one counterparty portfolio that contains 476 interest rate swaps, 36 interest rate swaptions and 223 interest rate caps/floors.

First, we compute the risk-free value $V^F = 2,737,702$ that is relatively straightforward as the risk-free portfolio value is what trading systems or pricing models normally report.

Second, we assume that there is counterparty credit risk but no collateral agreement. Based on the pricing model proposed by Xiao [2015], we compute the risky value of the portfolio as $V^N = 2,688,014$ after considering counterparty credit risk. By definition, the CVA without collateralization is equal to $CVA_N = V^F - V^N = 49,688$.

Next, we further assume that there is a CSA agreement in which the threshold is 2 million and the MTA is 100,000. The risky value of the portfolio is calculated as $V^C = 2,725,094$ according to Proposition 2. The CVA with collateralization is given by $CVA_C = V^F - V^C = 12,608$. Similarly, we can compute the CVAs under different collateral arrangements and present the results in Exhibit 9.

**Exhibit 9. Impact of Collateralization on CVA**

This exhibit shows that CVA increases with collateral threshold. The infinite collateral threshold is equivalent to no collateral agreement and the zero-value collateral threshold corresponds to full collateralization. An increase in collateral threshold leads to a decrease in collateralization.

| Effective Threshold | 0 | 2.1 Million | 4.1 Million | 6.1 Million | 8.1 Mil | Infinite ($\infty$) |
|---|---|---|---|---|---|---|
| CVA | 0 | 12,608 | 23,685 | 33,504 | 42,254 | 49,688 |

Exhibit 9 tells us that collateral posting can reduce CVA. Full collateralization makes a portfolio appear to be risk-free. An increase in collateral threshold leads to a rise in unsecured credit exposure, and thereby an increase in CVA. In particular, CVA reaches the maximum when the threshold is infinite representing no collateral arrangement.

We extend our analysis to other CSA portfolios. The results hold across different CSA counterparties and collateral agreements. Our findings show that collateral posting can reduce credit risk and CVA. The results also suggest a negative correlation between collateralization and CVA as an increase in collateralization causes a decrease in CVA charge and vice-versa. These findings improve our understanding of the relationship between collateralization and CVA.

**Impact of collateralization on market risk**

We study how collateralization impacts market risk by gauging VaR changes due to collateral arrangements. VaR is the regulatory measurement for assessing market risk. It is defined as the maximum loss likely to be suffered on a portfolio for a given probability defined as a confidence level over a given period time. In its most popular form, VaR measures 10-day $99^{th}$ percentile of potential loss that can be incurred.

There are three commonly used methodologies to calculate VaR – parametric, historical simulation and Monte Carlo simulation. Parametric model estimates VaR directly from the standard deviation of portfolio returns typically assuming returns are normally distributed. Historical simulation calculates VaR from the distribution of actual historical returns. Whilst Monte Carlo simulation computes VaR from a distribution constructed from random outcomes. In this paper, we calculate historical VaR.

For monitoring market risk, many organizations segment portfolios in some manner. They may do so by traders and trading desks. Typically, a market risk portfolio contains derivatives across multiple counterparties, while a counterparty portfolio comprises transactions among different traders and trading desks. We arbitrarily select a trading portfolio and then partition it into single-counterparty sub-portfolios. One sub-portfolio contains 172 interest rate swaps, 68 caps and floors, 25 European swaptions and 17 cancelable swaps.

First, we compute the VaR of the sub-portfolio without considering credit risk. The calculated result as of May 11, 2012 is -386,570. Here the VaR is a negative value by definition. However, the industry convention is to report VaR as a positive number that is the amount of money one can lose. Thus, people in the market usually say that the VaR is 386,570 in this case. This convention can be confusing in places. If one says that VaR increases, the numbers actually become smaller or move into the left tail.

For many reasons, both historical and practical, market and credit risk have often been treated as if they are unrelated source of risk: the risk types have been measured separately, managed separately, and economic capital against each risk type has been assessed separately.

However, market risk and credit risk actually reinforce each other. Default-induced credit losses can be driven by market price changes. At the same time, the changes in prices depend on the rating migration or increase in the default perception of the firm. The Basel Committee on Banking Supervision (see BCBS [2009]) also finds that those banks that were more severely affected in the global financial crisis measured their market and credit risk separately, whereas banks that used an integrated approach to market and credit risk measurement were much less impacted. Measuring market risk without considering credit risk may mask the significant impact of credit risk and often leads to underestimation of risk.

Therefore, we further calculate VaR by taking credit risk into account. Let us first assume that there is no CSA agreement. Each trade is valued by the risky model developed by Xiao [2015]. The VaR with credit risk is computed as -418,948. It can be seen that VaR has increased from 386,570 to 418,948 after accounting for credit risk. The above empirical results show that market fluctuation has a larger impact on VaR when credit risk is taken into account. Our research highlights the linkage between market and credit risk.

Next, we measure VaR again by considering both counterparty credit risk and collateralization. Assume that there is a CSA arrangement. The VaR values under different effective collateral thresholds are displayed in Exhibit 10.

**Exhibit 10. Impact of Collateralization on VaR**

This exhibit shows that VaR increases with collateral threshold. The infinite collateral threshold is equivalent to no collateral agreement and the zero-value collateral threshold corresponds to full collateralization.

| Effective Threshold | 0 | 2.1 Million | 4.1 Million | 6.1 Mil | Infinite ($\infty$) |
|---|---|---|---|---|---|
| VaR | -386,570 | -398,235 | -404,401 | -410,756 | -418,950 |

Exhibit 10 tells us that collateral posting actually can reduce VaR. The results show that market risk exposure rises as collateral threshold increases. In particular, VaR reaches the maximum when the threshold is infinite representing no collateral arrangement. We come to the same conclusion after repeating this test for other portfolios.

Our findings suggest that there are important interactions between market and credit risk. We find a positive correlation between market and credit risk as they increase or decrease together. We also find that collateralization is actually negatively correlated with market risk, i.e., an increase in collateralization causes a drop in market risk. Said differently, collateral posting can reduce market risk. Our research contributes to the understanding of the interaction between market and credit risk.

## Conclusion

This article addresses an important topic of the impact of collateralization on valuation and risk. We present a new model for pricing collateralized financial contracts based on the fundamental principal and legal structure of CSA. The model can back out market prices. This is very useful for pricing outstanding collateralized derivatives.

Empirically, we use a unique proprietary dataset to measure the effect of collateralization on pricing and compare it with model-implied prices. The empirical results show that the model-implied prices are quite close to the market-quoted prices, suggesting that the model is fairly accurate on pricing collateralized derivatives.

We find strong evidence that counterparty credit risk alone plays a significant but not overwhelming role in determining credit-related spreads. Only the joint effect of collateralization and credit risk has high explanatory power on unsecured credit costs. This finding suggests that failure to properly account for collateralization may result in significant mispricing of derivatives.

We also find evidence that there is a strong linkage between market and credit risk. Our research results suggest that banks and regulators need to think about an integrated framework to capture material interactions of these two types of risk. This requires all profits and losses are gauged in a consistent way across risk types as they tend to be driven by the same economic factors. Our finding leads to an improved understanding of the interaction between market and credit risk and how this interaction is related to risk measurement and management.

# Appendix

**Proof of Proposition 1:** The binomial default rule considers only two possible states: default or survival. For a discrete one-payment period $(t, T)$ economy, at time $T$, the contract either defaults with the default probability $q(t,T)$ or survives with the survival probability $p(t,T)$. The survival value is the promised payoff $X_T$ and the default payment is $C(T) + \varphi(T)(X_T - C(T))$. The value of this collateralized contract is the discounted expectation of all the payoffs and is given by

$$V^C(t) = E\{D(t,T)[q(t,T)(C(T) + \varphi(T)(X_T - C(T))) + p(t,T)X_T] | \mathcal{F}_t\} \tag{A1}$$

The collateral posted at time $t$ is defined in (5). The value of the collateral at time $T$ becomes $C(T) = C(t)/D(t,T)$, where we consider the time value of money only for collateral assets.

If $V^C(t) > H(t)$, we have $C(t) = V^C(t) - H(t)$ and

$$V^C(t) = V^N(t)/\bar{I}(t,T) - H(t)\bar{q}(t,T)(1 - \varphi(T))/\bar{I}(t,T) \tag{A2}$$

where $I(t,T) = D(t,T)[p(t,T) + \varphi(T)q(t,T)]$, $\bar{I}(t,T) = E(I(t,T)|\mathcal{F}_t)$, and $V^N(t) = E[I(t,T)X_T|\mathcal{F}_t]$.

In this case, $V^C(t) > H(t)$ is equivalent to $V^N(t) > H(t)$.

If $V^C(t) \leq H(t)$, we have $C(t) = 0$ and

$$V^C(t) = V^N(t) \leq H(t) \tag{A3}$$

Combining the two cases of $V^C(t) > H(t)$ and $V^C(t) \leq H(t)$, we get

$$V^C(t) = 1_{V^N(t) \leq H(t)} V^N(t) + 1_{V^N(t) > H(t)} \left[ V^N(t)/\bar{I}(t,T) - H(t)\bar{q}(t,T)(1-\varphi(T))/\bar{I}(t,T) \right] \quad \text{(A4)}$$

or

$$V^C(t) = E[F(t,T)X_T | \mathcal{F}_t] - G(t,T) \quad \text{(A5a)}$$

where

$$F(t,T) = \left( 1_{V^N(t) \leq H(t)} + 1_{V^N(t) > H(t)} / \bar{I}(t,T) \right) I(t,T) D(t,T) \quad \text{(A5b)}$$

$$G(t,T) = 1_{V^N(t) > H(t)} H(t) \bar{q}(t,T)(1-\varphi(T))/\bar{I}(t,T) \quad \text{(A5c)}$$

**Proof of Proposition 2:** Let $t = T_0$. On the first payment day $T_1$, let $V^C(T_1)$ denote the CSA value of the contract excluding the current cash flow $X_1$. According to Proposition 1, the CSA value of the contract at $t$ is given by

$$V^C(t) = E[F(T_0, T_1)(X_1 + V^C(T_1)) | \mathcal{F}_t] - G(T_0, T_1) \quad \text{(A6)}$$

Similarly, we have

$$V^C(T_1) = E[F(T_1, T_2)(X_2 + V^C(T_2)) | \mathcal{F}_{T_1}] - G(T_1, T_2) \quad \text{(A7)}$$

Note that $F(T_0, T_1)$ and $G(T_0, T_1)$ are $\mathcal{F}_{T_1}$-measurable. According to the *taking out what is known* and *tower* properties of conditional expectation, we have

$$\begin{aligned} V^C(t) &= E[F(T_0, T_1)(X_1 + V^C(T_1)) | \mathcal{F}_t] - G(T_0, T_1) \\ &= \sum_{i=1}^{2} E\left[ \prod_{j=0}^{i-1} (F(T_j, T_{j+1})) X_i \Big| \mathcal{F}_t \right] + E\left[ \prod_{j=0}^{1} (F(T_j, T_{j+1})) V^C(T_2) \Big| \mathcal{F}_t \right] \\ &\quad - \sum_{i=0}^{1} E\left[ \prod_{j=0}^{i-1} (F(T_j, T_{j+1})) G(T_i, T_{i+1}) \Big| \mathcal{F}_t \right] \end{aligned} \quad \text{(A8)}$$

By recursively deriving from $T_2$ forward over $T_m$, where $V^C(T_m) = X_m$, we obtain

$$V^C(t) = \sum_{i=1}^{m} E\left[ \prod_{j=0}^{i-1} (F(T_j, T_{j+1})) X_i \Big| \mathcal{F}_t \right] - \sum_{i=0}^{m-1} E\left[ \prod_{j=0}^{i-1} (F(T_j, T_{j+1})) G(T_i, T_{i+1}) \Big| \mathcal{F}_t \right] \quad \text{(A9)}$$

where $\prod_{j=0}^{i-1} F(T_j, T_{j+1}) = 1$ is an empty product when $i = 0$.